\documentstyle[preprint,aps]{revtex}
\begin{document}
\preprint{IMSc-95/21}
\title{A Mechanism for Instanton induced Chiral Symmetry Breaking in
QCD}
\author{Rahul Basu and R. Parthasarathy \cite{email}}
\address{The Institute of Mathematical Sciences, Madras-600 113, INDIA}

\maketitle
\begin{abstract}
We propose a mechanism for instanton induced chiral symmetry breaking in QCD
with fundamental scalars. The model Lagragian that
we use has the same symmetry properties as QCD. The scalar fields develop
vacuum
expectation values at a non-trivial minimum and generate masses for
thhe light quarks. The minimization condition is also used to break the
$SU(N_f)$ flavour symmetry in order to make the $s$ quark heavier than the
two lighter ones. Thus a vacuum of the theory that is not chirally invariant
is obtained.
\end{abstract}


\section{Introduction}

In the gauge theory of strong interactions viz. QCD, the mechanism of chiral
symmetry breaking is an important issue. The current quark
masses for u, d and s quarks are known not to be zero. In the absence of
explicit mass terms for the quarks, the Lagrangian for QCD has exact chiral
symmetry $SU(N_f)_L\times SU(N_f)_R$, in addition to $U(1)_V$ (baryon number
conservation) and $U(1)_A$ which, as is well known, is violated by quantum
effects. Apart from the issue of the non-trivial nature of the QCD vacuum
with the quark condensate $<0|{\bar q}q|0>\neq 0$, the nature of the chiral
phase
transition in QCD continues to be of interest. This has been studied in
various ways, by assuming an effective theory with the desired properties.
In particular, we would like to mention here the work of Pisarski and
Wiilczek \cite{pw} and the related earlier work of Raby \cite{raby}. While
the former authors studied the restoration of chiral symmetry at finite
temperature, the latter study concentrated more on phenomenological aspects
of the $\eta$ decay width and related issues. However, the usual practise
followed by these authors to break the symmetry in the first place is
to add a term $c(detM + detM^\dagger)$, where $M$ is a colour singlet complex
$N_f\times N_f$ matrix, which is also used to give mass to the $\eta^\prime$
meson.

The effect of instantons in QCD (without fundamental scalars) to produce
chiral symmetry breaking is by now an old isuue  and was studied amongst
others, by Callan, Dashen and Gross \cite{cdg}, by 't Hooft \cite{thooft}
in the context of the $U(1)$ problem and by Shuryak \cite{shuryak}.
However, the mechanism of dynamical chiral
symmetry breaking continues to be an important problem in QCD.

In this paper, we re-examine the issue in the context of the effect of
instantons in QCD with fundamental scalars described by a color singlet
$N_f\times N_f$ complex matrix, allowing these scalars to have only a
kinetic energy term and Yukawa coupling with the quark fields.  The idea of
treating scalars as elementary fields interacting with quarks has been
earlier considered by D'yakonov and Eldes \cite{de}, Dhar, Shankar and Wadia
\cite{dsw} and others in the Nambu-Jona Lasinio model motivated effective
QCD Lagrangian. However our model differs from these in the sense the we
evaluate the effective potential for $M$ by integrating out the quark fields
in the instanton background at the one-loop level. In that sense we use a
fundamental rather than an effective Lagrangian. A similar approach to
explain small fermion masses in QCD has been considered by Huang and
Viswanathan \cite{hv} with $M$ as a color singlet complex {\em single
component} scalar field. We show that with $M$ as a $N_f\times N_f$
complex matrix, the effective potental induced by the instanton one-loop
effects leads to chiral symmetry breaking by producing a $(detM +
detM^\dagger)$ from the zero modes of the Dirac operator.  However before we
go on to describe the work in the rest of the paper, it is appropriate at
this point to motivate the use of our Lagrangian which we shall be using
subsequently.

Though QCD is and has been an acceptable theory of strong interactions, its
low energy behavior still eludes explicit calculations, where effective
theories like the Gell-Mann Levy linear $\sigma$ model are used to give a
field theoretical description of the strong interactions. The linear sigma
model is not asymptotically
free, and therefore, inspite of being perturbatively renormalizable, is
usually considered a low energy effective theory. Recently however, the chiral
sigma model with quarks substituted for nucleons \cite{krs}  has found
support as a reasonable description of the nucleon \cite{bb}, strong
interaction properties at finite temperature and baryon density
\cite{goksch,gottlieb,soni1} even at scales well above chiral
restoration,  as well as weak interaction properties
\cite{partha}. At first sight this is unexpected from an effective theory.
However it has very recently been shown that the chiral linear sigma model
with quarks, when coupled to gluons can be asymptotically free \cite{soni3}.
Lattice studies \cite{goksch,gottlieb} indicate that the chiral sigma model
with quarks reproduce QCD lattice results rather well at finite temperature
with the pions and sigma being elementary for all $T >  T_\chi$ except not
as Goldstone bosons. However gluons were neglected in these references. In
this paper, we consider QCD with the addition of a color singlet scalar
multiplet as fundamental fields.  We will allow for $SU(N_f)\times SU(N_f)$
flavor symmetry which entails taking $\sigma=\sigma^a\lambda^a,
\pi=\pi^a\lambda^a$, collectively denoted by a complex $N_f\times N_f$
matrix $M$ as stated earlier ($\lambda^a$ are the usual Gell-Mann $SU(3)$
matrices for $N_f =3$).  We use this as a candidate theory for strong
interactions with quarks, gluons and mesons as elementary excitations. We
assume that we are in a phase of the theory (dictated by  initial conditions
on the QCD and Yukawa coupling and the renormalization scale $\mu$) where
not only the quarks but also the scalar and pseudoscalar mesons contained in
$M$ are elemenatary.

We describe the model in Section II and evaluate the quantum one-loop
effective action with instanton as the classical background. In Section III
the effective potential is obtained in the dilute gas approximation for the
instantons. The potential is examined for a non-trivial minimum in Section
IV and the quark mass generation is studied. Improvements following
\cite{pw} and \cite{raby} are also examined. The results are summarized in
Sec. V.

\section{The model}

The proposed Lagrangian consists of the usual Lagrangian of QCD and an
$(N_f\times N_f)$ color singlet complex matrix $M$ whose matrix elements
consist of elementary scalars and pseudoscalars, interacting with quarks
through the Yukawa coupling, viz.
\begin{eqnarray}
{\cal L}& =& - {1\over 4}F^a_{\mu\nu}F^{\mu\nu a} + {\bar\psi}iD\!\!\!\!/\psi
+ i\theta {\bar F}F \nonumber \\
&& + g_y({\bar\psi}_LM\psi_R+{\bar\psi}_RM^\dagger \psi_L) +
{1\over 2} \partial_\mu M\partial^\mu M^\dagger , \label{lag}
\end{eqnarray}
where $D\!\!\!\!/ = \partial\!\!\!\!/ -igA\!\!\!\!/$,\  $g_y$ is the
Yukawa coupling and
we have added a kinetic energy term for M. A few remarks about (\ref{lag})
are in order. This Lagrangian possesses global $SU(N_f)_L\times SU(N_f)_R$
chiral invariance and of course, $SU(3)_c$ local color invariance. In
addition it has $U(1)_V$ (baryon number) and $U(1)_A$ invariances, the
latter being broken by quantum corrections, leaving only a discrete
$Z(N_f)_A$ symmetry. The raison d'etre for using this Lagrangian has already
been explained in the Introduction. The topological term $i\theta{\bar F}F$
has been added although we know that experimental limits on the neutron
dipole moment sets a limit on $\theta < 10^{-9}$.

The partition function for (\ref{lag}) is
\begin{equation}
Z=\int {\cal D}\bar\psi{\cal D}\psi{\cal D}A_\mu{\cal D}M{\cal D}M^\dagger
  e^{i\int d^4x{\cal L}[\psi,\bar\psi,A_\mu,M,M^\dagger]}
\end{equation}

We evaluate the partition function, first by going over to Euclidean space
and choosing instanton background for the gluon field. This consists in
writing $A_\mu^a = \bar A^a_\mu (instanton)+a^a_\mu$. The integration over
$\psi$ and $\bar\psi$ is done first, giving the result

\begin{eqnarray}
Z&=&\int {\cal D}A_\mu{\cal D}M{\cal D}M^\dagger e^{-\int d^4x[- {1\over
4}F^a_{\mu\nu}F^{\mu\nu a} +i\theta{\bar F}F]} \nonumber \\
&& det(i{\bar D}\!\!\!\!/+ReM + i\gamma_5 Im M),
\end{eqnarray}
where ${\bar D}\!\!\!\!/=\partial\!\!\!\!/-ig\bar A\!\!\!\!/$ (instanton)and
the
determinant above is over Lorentz and flavour indices. The functional
integration over $A_\mu^a$ can be split over instanton locations $\bar
A^a_\mu$(instanton)and and an integral over the fluctuations $a^a_\mu$
around the instantons. We make the usual dilute gas approximation for the
instantons. The Gaussian integration over $a^a_\mu$ givs a contribution
\cite{thooft} which we denote by $K$ and the integration over $\bar A^a_\mu$
(instanton) is replaced by a summation over instanton winding numbers, apart
from an overall Jacobian for the change of measure.The determinant
$det(iD\!\!\!\!/+ReM+i\gamma_5 Im M)$ gives a contribution $det M$ and $det
M^\dagger$ for each zero mode of the $D\!\!\!\!/$ operator
$(\gamma_5\psi_0=\pm\psi_0)$ and $det^{\prime}(-D\!\!\!\!/^2 + MM^\dagger)$
from the non-zero  modes.  The result of all this is

\begin{eqnarray}
Z &=&\int{\cal D}M{\cal D}M^\dagger\sum_{n_+}\frac{1}{n_+!}e^{in_+\theta}
     K^{n_+}(det M)^{n_+} \nonumber \\
&&     \sum_{n_-}\frac{1}{n_-!}e^{in_-\theta}
	  K^{n_-}(det M)^{n_-}    \nonumber \\
&& det^{'}(-D\!\!\!\!/^2 + MM^\dagger)e^{-\int d^4x\partial_\mu M\partial^\mu
M^\dagger}   \label{part-func}
\end{eqnarray}
where we have taken into account the exchange symmetry of the instantons.
The summations over $n_+$ and $n_-$ can be carried out and the effective
action $\Gamma$ defined through $Z=\int {\cal D}M{\cal D}M^\dagger
e^{-\int \Gamma}$, can be written down to be

\begin{eqnarray}
\Gamma &=& \frac{1}{2}\partial_\mu M\partial M^\dagger + Kdet M + Kdet
M^\dagger \nonumber \\
&& + ln det(-D\!\!\!\!/^2+MM^\dagger) \label{eff-act}
\end{eqnarray}
where we have taken, without loss of generality $\theta = 0$.

\section{One Loop Effective Potential}

The result of the previous section shows that with the model proposed, the
effect of the quantum one loop calculations around an instanton background
is to produce an effective action given in (\ref{eff-act}). Before
continuing with our study, it will be instructive to compare our effective
action (\ref{eff-act}) with that of Raby \cite{raby} and Pisarski and
Wilczek \cite{pw}. The point we wish to make is that the term $(det M + det
M^\dagger)$ that they add for phenomenological reasons is derived in our
paper as a contribution arising out of the zero modes of the Dirac operator
in the instanton background. The importance of this term, as pointed out in
these references, is to break $U(N_f)\times U(N_f) \rightarrow SU(N_f)\times
SU(N_f)\times U(1)_V$ which is the quantum symmetry of QCD. To be precise, the
Lagrangian ({\ref{lag}) has $G=SU(3)_c\times SU(N_f)\times SU(N_f)\times U(1)_V
\times U(1)_A$ symmetries with the generators of G acting on $\psi$. By
integrating out the fermion field, we obtain an effective action
(\ref{eff-act}) for $M$ which has the same symmetry $G$ with the generators now
acting on $M$. This  term makes the $\eta^\prime$ massive as the broken
$Z_A(N_f)$ symmetry is
discrete. In order to appreciate the role of this term in our model, we see
from (\ref{part-func}) and (\ref{eff-act}) that if $\theta\neq 0$, then we
would have gotten $Kcos\theta(detM+detM^\dagger)+Ki sin\theta(detM-det
M^\dagger)$. The second term violates CP and hence we have put $\theta=0$
consistent with the data on the electric dipole moment of the neutron.

The other important term is expected to come from the contribution of the
non-zero eigenvalue sector and as shown, contributes $ln
det^\prime(-\bar D\!\!\!\!/^2 + MM^\dagger)$. This does not break $U(1)_A$ and
will be a function of $MM^\dagger$. The exact calculation of this piece if
of course difficult. Following 't Hooft \cite{thooft} we see that the
regularized product of the non-vanishing eigenvalues is proportional to the one
with $\bar A_\mu = 0$, the proportionality constant being dependent on the
instanton quantum numbers (see, for example, Eq. 6.15 of \cite{thooft}).
Therefore, we write this term as

\begin{equation}
A \ ln det^\prime(-k^2+MM^\dagger) \nonumber
\end{equation}
where $A$ contains the effect of instantons taken to be compact in each small
volume $\Delta V$ of space-time and the ghost determinant arising from gauge
fixing. We further consider a basis in
which the complex matrix $M$  is diagonal with elements $\lambda_i,
(i=1,\ldots,N_f)$. This is done by considering the term
$({\bar\psi}_LM\psi_R+{\bar\psi}_RM^\dagger \psi_L)$ in the Lagrangian. We
diagonalize $M$,the complex $N_f\times N_f$ matrix, by independent left and
right unitary transformations viz. $U^{\dagger}_LMU_R=\Lambda$ where
$\Lambda_{ij}=\lambda_i\delta_{ij}$. This transformation, of course, introduces
a mixing of flavours in $\psi$; nevertheless, the strong interaction part
$\bar\psi A\!\!\!/\psi (=\bar\psi_L A\!\!\!\!/\psi_L+\bar\psi_R
A\!\!\!\!/\psi_R)$ is unaffected by this mixing. In general these
$\lambda_i$'s are complex.
Using $\epsilon$-regularizing scheme (see, for example, \cite{ss}) we can then
write

\begin{equation}
A\ ln det^\prime(-k^2+MM^\dagger)\simeq A\sum_{i=1}^{N_f}\vert \lambda_i\vert^4
ln(\frac{\vert\lambda_i\vert^2}{\mu^2})
\end{equation}
where $\mu$ is a regularization scale.

Combining all this, the effective potential for $M$ becomes

\begin{equation}
V_{eff}=K\prod_{i=1}^{N_f}\lambda_i + K\prod_{i=1}^{N_f}\lambda_i^*
+ A\sum_{i=1}^{N_f}\vert \lambda_i\vert^4
ln(\frac{\vert\lambda_i\vert^2}{\mu^2})
\label{eff-pot}
\end{equation}
Note that the first two terms are a consequence of the topologically
non-trivial instanton background for gluons that we have chosen. Such terms
will not be present in a usual topological trivial configuration like, for
instance, the Saviddy background. This will prove to be crucial for our
purposes.

\section{Mechanism of Chiral Symmetry Breaking}

The minimum of the effective potential (\ref{eff-pot}) will provide us
with the vacuum expectation values for the diagonal elements of M. To
simplify the algebra, we will henceforth restrict ourselves to $N_f=3$.
We are interested in the minimum of the effective potential (\ref{eff-pot})
so that the vacuum expectation values $<M>$ can be identified with
$\lambda_i, (i=1,2,3)$. The minimisation with respect to
$\lambda_1,\lambda_2, \lambda_3$ yields

\begin{eqnarray}
K\lambda_2\lambda_3&=&-A\lambda_1^*\vert\lambda_1\vert^2
(1+2ln(\frac{\vert\lambda_1\vert^2}{\mu^2})) \nonumber \\
K\lambda_3\lambda_1&=&-A\lambda_2^*\vert\lambda_2\vert^2
(1+2ln(\frac{\vert\lambda_2\vert^2}{\mu^2})) \nonumber \\
K\lambda_1\lambda_2&=&-A\lambda_3^*\vert\lambda_3\vert^2
(1+2ln(\frac{\vert\lambda_3\vert^2}{\mu^2})) \nonumber \\
\end{eqnarray}
from which, it immediately follows
\begin{eqnarray}
\vert\lambda_1\vert^4(1+2ln(\frac{\vert\lambda_1\vert^2}{\mu^2}))
& =&\vert\lambda_2\vert^4(1+2ln(\frac{\vert\lambda_2\vert^2}{\mu^2}))
\nonumber \\
&=& \vert\lambda_3\vert^4(1+2ln(\frac{\vert\lambda_3\vert^2}{\mu^2}))
\nonumber \\
   \label{min}
\end{eqnarray}

At this stage, the $\lambda_i$'s are constants representing vacuum
expectation values $<M>$. In general, the $\lambda_i$'s are complex. Since
however these are now space-time independent constants, their phases can be
absorbed in $U_L$ or $U_R$. With this, the $\lambda_i$'s can be treated as
real.

The trivial solution to (\ref{min}) is of course $\lambda_1=\lambda_2=
\lambda_3 =0$ which corresponds to
$V_{eff}=0$. However, a non-trivial solution to (\ref{min}) is given by
\begin{equation}
\lambda_1=\lambda_2= \lambda_3 =\lambda \neq 0
\label{sol}
\end{equation}
where $\lambda$ is given by
\begin{equation}
\lambda=\mu e^{-(1+{K\over A\lambda})/4}
\end{equation}

The value of the effective potential for this solution is

\begin{equation}
V_{eff}(\lambda_1=\lambda_2=\lambda_3 =\lambda)=-2A\lambda^4(1+
ln(\frac{\lambda}{\mu}))
\end{equation}

We require this to be lower than $V_{eff}=0$ (corresponding to the trivial
solution $\lambda_1=\lambda_2=\lambda_3 =0$) which is possible if

\begin{equation}
\mu < e\lambda
\end{equation}

This sets the energy scale $\mu$ for which chiral symmetry breaking is
possible. The symmetric solution (\ref{sol})spontaneously breaks
$SU(3)_L\times SU(3)_R$ to $SU(3)$ symmetry.
We stress again that the $(detM + detM^\dagger)$
coming from the instanton zero modes is crucial for this symmetry
breaking. It is also clear that this breaking gives rise (in the 3 flavor
case) to masses for the quarks

\begin{equation}
m_u=m_d=m_s=g_y\lambda. \label{mass}
\end{equation}
Thus as conjectured in early works (see, for example, \cite{cdg}), this model
provides us with an explicit realization of chiral symmetry through
instantons, due to the presence of fundamental scalars. A numerical value of
$\mu$ cannot be obtained since $\lambda$ is not specified in terms of known
masses. However an estimate can be made if the Yukawa coupling in
(\ref{mass}) is taken to be approximately around unity. Then $\lambda$ can
be identified with the quark masses as given in the equation above. Taking
$m_s \sim 150$ MeV, we get $\mu < 300$ MeV, which sets the scale in this
model for chiral symmetry breaking.

The minimization condition also provides us with a more phenomenologically
realistic scenario where all the quark masses are not equal. For this, we will
assume that $\lambda_1=\lambda_2=\lambda\neq \lambda_3$. In particular we look
for a solution for which $\lambda_3=a\lambda$, such that
\begin{equation}
m_u=m_d\neq m_s  \label{broken}
\end{equation}
Substituting in the minimization condition
(\ref{min}), the result $\lambda_3=a\lambda$ we get the following equation
\begin{equation}
\frac{a^4}{1-a^4}ln a^2 = {1\over 2} + ln(\frac{\lambda^2}{\mu^2}).
\end{equation}
This equation is numerically solved for many values of $a$
(see Appendix). In
particular we choose a value of $a=20$ as the ratio of the $s$ current
quark mass to the $u$ (or $d$) current quark mass. This immediately gives us
a value of $\mu \simeq 26\lambda$ which for $g_y \sim 1$ gives $\mu\simeq 210
$MeV, consistent with our previous estimate $\mu < 300$ MeV.
We have broken the degeneracy between the quark masses between the $s$
quark and the two lighter quarks, thereby breaking the $SU(3)$ flavour
symmetry to isospin ($SU(2)$) and strangeness ($U(1)$).  This kind of
sequential symmetry breaking was envisaged long ago by Gell-Mann \cite{mgm} by
introducing $(\bar\psi_L M\psi_R + \bar\psi_R M^{\dagger}\psi_L)$ as an
ideal pattern of couplings to ``mesons''. He added two types of terms; one
is $u_0 \sim -m\bar\psi\psi$ which corresponds to, in our case
$\lambda_1=\lambda_2=\lambda_3=\lambda$ (Eq. \ref{mass}) and a term $u_8\sim
m_{02}(\bar NN-2\bar\lambda \lambda)$ (in the notation of \cite{mgm}) which
corresponds to $\lambda_1=\lambda_2=\lambda; \lambda_3\neq \lambda$,
corresponding, in our case to (\ref{broken}). In this way, the effect of the
$u_0$ and
$u_8$ terms of Gell-Mann,  which have been introduced later by many other
authors (see, for example, \cite{raby}), can be reproduced from the solution
(\ref{min}) of the minimum of the effective potential  in our model. With
the addition of these terms, the study of the effective action
(\ref{eff-act}) proceeds along the lines elucidated by Raby, by expanding
$<M>$ around $<M>_0$.

\section{Conclusion}

A mechanism for producing chiral symmetry breaking in QCD with scalars,
is demonstrated. The model Lagrangian has the
same symmetry properties as QCD. The effective potential for the ``field''
$M$ is obtained in an instanton background and is shown to have a non
trivial minimum. This generates masses for the quarks due to the
breakdown of chiral symmetry. The connection between this study and earlier
work {\it viz.} that of Pisarski and Wilczek and of Raby is pointed out. The
method of Raby of introducing tadpole terms to break the $SU(3)$ flavor
symmetry is realized here by the minimization condition with $m_u=m_d\neq
m_s$.
Thus a vacuum of the theory that is not
chirally invariant (as dictated by QCD sum rules and other studies)
has been obtained.

\section{Appendix}

To see that equation (\ref{min}) can accomodate many  solutions
where the $\lambda$'s are not degenerate, we make the change of variable
$x=\frac{\lambda_1^4}{\mu^4}$ and $y=\frac{\lambda_3^4}{\mu^4}$. Then the
minimization condition (\ref{min}) becomes
\begin{equation}
ln (e^x x^x)  = ln (e^y y^y)
\end{equation}
Now taking $y$ to be a multiple of $x$, say $y=bx$ (where $b=a^4$ as defined
earlier in the text), the above equation reduces to
\begin{equation}
e^{x(1-b)}x^{x(1-b)}b^{-bx}=1
\end{equation}

The easiest way to solve this equation is graphically by plotting the
function $f(x)=e^{x(1-b)}x^{x(1-b)}b^{-bx}$ as a function of $x$ for
different values of the parameter $b$ and checking where it cuts the $f(x)=1$
line. This is shown graphically in the figure. Notice that we have to
choose very large values of $b=a^4$ since $a=20$ from phenomenological
considerations as mentioned before. Note also that the $f(x)=1$ line also
coincides with the $b=1$ case when $f(x)=1$ for all values of $x$. We have
used different values of $b$ to indicate that they all cut the $f(x)=1$ line
at some point corresponding to some value of $x$ which, in turn, will
provide us with the $\lambda$ to $\mu$ ratio, as indicated in section IV.

\acknowledgements

We would like to thank H. S. Sharatchandra for many discussions and several
illuminating comments. We would also like to thank  R. Anishetty,
R. Ramachandran and G. Rajasekaran for discussions.

\vskip 1cm
{\bf \Large Figure caption}
\vskip 0.5cm
\noindent Plot of $f(x)=e^{x(1-b)}x^{x(1-b)}b^{-bx}$ as a function of $x$ for
different $b$'s. The horizontal line is the $f(x)=1$ line (as also the $b=1$
line for all x).


\begin{references}
\bibitem[\ddag]{email}{\em electronic address:}rahul,sarathy@imsc.ernet.in
\bibitem{pw}
R. D. Pisarski and F. Wilczek, Phys. Rev. {\bf D29}, 338 (1984).
\bibitem{raby}
S. Raby, Phys. Rev. {\bf D13}, 2594 (1976).
\bibitem{cdg}
C. G. Callan Jr., R. F. Dashen and D. J. Gross, Phys. Rev. {\bf
D17} 2717 (1978).
\bibitem{thooft}
G. 't Hooft, Phys. Rev. {\bf D14}, 3432 (1976).
\bibitem{shuryak}
E. V. Shuryak,  Nucl. Phys. {\bf B203} 93 (1982);
Nucl. Phys.{\bf B214} 237 (1983).
\bibitem{de}
D. I. D'yakonov and M. I. Eldes, JETP Letters, {\bf 38}, 358 (1983).
\bibitem{dsw}
A. Dhar, R. Shankar and S. R. Wadia, Phys. Rev. {\bf D31} 3256 (1985).
\bibitem{hv}
Z. Huang and K. S. Vishwanathan, Z. Phys. {\bf C55}, 171(1992)
\bibitem{krs}
S. Kahane, G. Ripka and V. Soni, Nucl Phys. {\bf A415}, 351
(1984).
\bibitem{bb}
M. C. Birse and M. K. Banerjee, Phys. Lett. {\bf B136},
284(1984), Phys. Rev. {\bf D31}, 1118 (1985)
\bibitem{gottlieb}
S. Gottlieb, et al., {\em Thermodynamics of Lattice QCD
with two light flavours on a $16^3 \times 8$ lattice}, ANL preprint no.
ANL-HEP-PR-92-57.
\bibitem{goksch}
A. Goksch, Phys. Lett. {\bf  B152}, 1701 (1991).
\bibitem{soni1}
V. Soni, Phys. Lett. {\bf B152}, 231 (1985).
\bibitem{partha}
R. Parthasarathy, Invited talk given at CAP Summer Institute,  Queens
University, Kingston, Canada, July 1993.
\bibitem{soni3}
V. Soni, NPL preprint, 1995 (hep-ph/9505290).
\bibitem{ss}
A. Salam and J. Strathdee,  Nucl. Phys. {\bf B90},203 (1975)
\bibitem{mgm}
M. Gell-Mann, Lecture notes of the TIFR summer school in Theoretical
Physics, Bangalore, India,  1961.
\end{references}
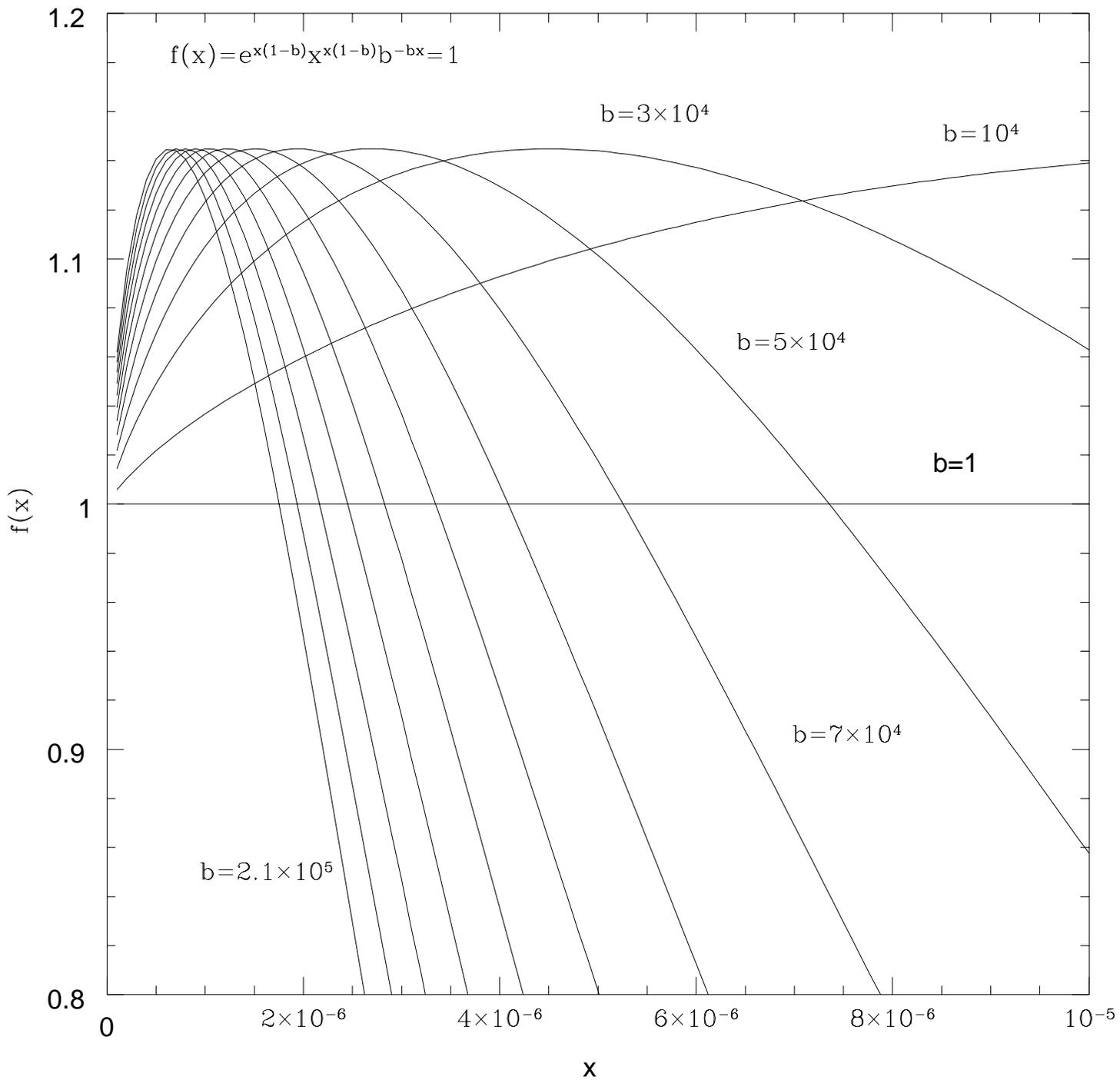
\end{document}